\documentclass[twocolumn,superscriptaddress,longbibliography]{revtex4-2}
\usepackage{graphicx}
\usepackage{epstopdf}
\usepackage{amsmath}
\usepackage{color}
\usepackage{makecell}
\usepackage{float}
\usepackage{fancyhdr}
\usepackage{bm}

\begin{document}
\pagestyle{fancy}
\fancyhf{}              
\fancyfoot[C]{\thepage} 
\renewcommand{\headrulewidth}{0pt}
\renewcommand{\footrulewidth}{0pt}

	\title{Intertwined spin-charge stripe order and polar lattice distortion in La$_{3}$Ni$_{2}$O$_{7}$}
	\author{Xiaoying Li}
\affiliation{Key Laboratory of Materials Physics, Institute of Solid State Physics, HFIPS, Chinese Academy of Sciences, Hefei 230031, China}
\affiliation{Science Island Branch of Graduate School, University of Science and Technology of China, Hefei 230026, China}

\author{Wenqian Tu}
\affiliation{Key Laboratory of Materials Physics, Institute of Solid State Physics, HFIPS, Chinese Academy of Sciences, Hefei 230031, China}
\affiliation{Science Island Branch of Graduate School, University of Science and Technology of China, Hefei 230026, China}

\author{Run Lv}
\affiliation{Key Laboratory of Materials Physics, Institute of Solid State Physics, HFIPS, Chinese Academy of Sciences, Hefei 230031, China}
\affiliation{Science Island Branch of Graduate School, University of Science and Technology of China, Hefei 230026, China}

\author{Li'e Liu}
\affiliation{Key Laboratory of Materials Physics, Institute of Solid State Physics, HFIPS, Chinese Academy of Sciences, Hefei 230031, China}
\affiliation{Science Island Branch of Graduate School, University of Science and Technology of China, Hefei 230026, China}

\author{Dingfu Shao}
\affiliation{Key Laboratory of Materials Physics, Institute of Solid State Physics, HFIPS, Chinese Academy of Sciences, Hefei 230031, China}

\author{Yuping Sun}
\affiliation{Anhui Province Key Laboratory of Low-Energy Quantum Materials and Devices, High Magnetic Field Laboratory, HFIPS, Chinese Academy of Sciences, Hefei 230031, China}
\affiliation{Key Laboratory of Materials Physics, Institute of Solid State Physics, HFIPS, Chinese Academy of Sciences, Hefei 230031, China}
\affiliation{Collaborative Innovation Center of Microstructures, Nanjing University, Nanjing 210093, China}

\author{Wenjian Lu}
\thanks{Corresponding author: wjlu@issp.ac.cn}

\affiliation{Key Laboratory of Materials Physics, Institute of Solid State Physics, HFIPS, Chinese Academy of Sciences, Hefei 230031, China}
	
	\begin{abstract}		
		The low-temperature density-wave state of La$_3$Ni$_2$O$_7$ hosts pronounced spin-density-wave (SDW) order, while recent experiments further reveal charge redistribution and a concomitant lattice-symmetry lowering. However, the microscopic relationship among spin, charge, and lattice remains unclear. Using first-principles calculations, we investigate the pressure evolution of the electronic structure and static spin susceptibility of La$_3$Ni$_2$O$_7$, together with the energetics and lattice response of representative magnetic configurations. We trace the SDW instability to strong Fermi-surface nesting and find that the high-pressure spin response closely tracks $T_{\mathrm C}$, suggesting spin-fluctuation-mediated pairing. Among the candidate magnetic states considered, the spin-charge-stripe states emerge as energetically favored and dynamically stable, developing pronounced disproportionation of both the local Ni moments and the Ni--O bond lengths. Remarkably, the lowest-energy $\bm{a}$-stripe state spontaneously relaxes into the experimentally proposed polar Am2m structure through a polar distortion along the $\bm{b}$ axis. These results establish a unified picture in which spin, charge, and lattice responses are strongly intertwined in the low-pressure density-wave state, while spin fluctuations remain a plausible ingredient of superconductivity under pressure.		
	\end{abstract}
	
	\maketitle
	
	\section{Introduction}
	
   Uncovering the mechanism of high-temperature superconductivity remains a central challenge in condensed-matter physics. The recent discovery of superconductivity near 80 K in the bilayer Ruddlesden--Popper (RP) nickelate La$_3$Ni$_2$O$_7$ under high pressure \cite{Sun2023} has opened a new platform for exploring unconventional superconductivity beyond the cuprates and iron-based superconductors. In such systems, the interplay among spin, charge, and lattice degrees of freedom can give rise to competing or intertwined ordered states, exemplified by the charge-density-wave (CDW) order in cuprates and the spin-density-wave (SDW) order in iron-based superconductors \cite{Chang2012,Kontani2021}. Mounting experimental evidence likewise identifies SDW order as the dominant instability of the low-temperature density-wave phase in La$_3$Ni$_2$O$_7$, whereas the existence and microscopic role of concomitant charge modulation remain unsettled \cite{Liu2024NC,Wu2025PRB}.

   A broad range of experimental probes, including NMR \cite{Kakoi2024,Zhao2025}, NQR \cite{Luo2025}, RIXS \cite{Chen2024NC}, RSXS \cite{Gupta2025}, and $\mu$SR \cite{Chen2024PRL}, provides compelling evidence for pronounced SDW order in the low-temperature phase of La$_3$Ni$_2$O$_7$ at ambient and low pressures. In particular, RSXS has directly resolved unidirectional diagonal double-stripe magnetic order with a propagation vector close to $\bm{Q}_{\mathrm{SDW}}\approx(\pi/2,\pm\pi/2)$, consistent with antiferromagnetic interlayer coupling. Previous theoretical studies based on tight-binding models, the random-phase approximation (RPA), and DFT+DMFT have connected the magnetic instability of La$_3$Ni$_2$O$_7$ to its underlying electronic structure \cite{Luo2023,Hu2025,Lu2025,Christiansson2023}. More recently, first-principles-based RPA calculations reproduced an SDW instability close to the measured wave vector and attributed it to Fermi-surface (FS) nesting \cite{Liu2025PRB}. However, these studies have largely treated the electronic and magnetic instabilities within predetermined structural frameworks, leaving the possible feedback among spin, charge, and lattice degrees of freedom insufficiently explored.

    Recent high-precision synchrotron X-ray diffraction (XRD) has further suggested that La$_3$Ni$_2$O$_7$ may adopt a polar Am2m structure at ambient pressure \cite{Misawa2026}. This lower-symmetry structure contains inequivalent Ni sites and pronounced Ni--O bond disproportionation, features consistent with checkerboard-like charge ordering. These observations raise a key question: does the intrinsic coupling among spin, charge, and lattice degrees of freedom play an essential role in stabilizing the low-temperature SDW state?    
    
    \begin{figure*}
    	\centering
    	\includegraphics[width=0.9\linewidth]{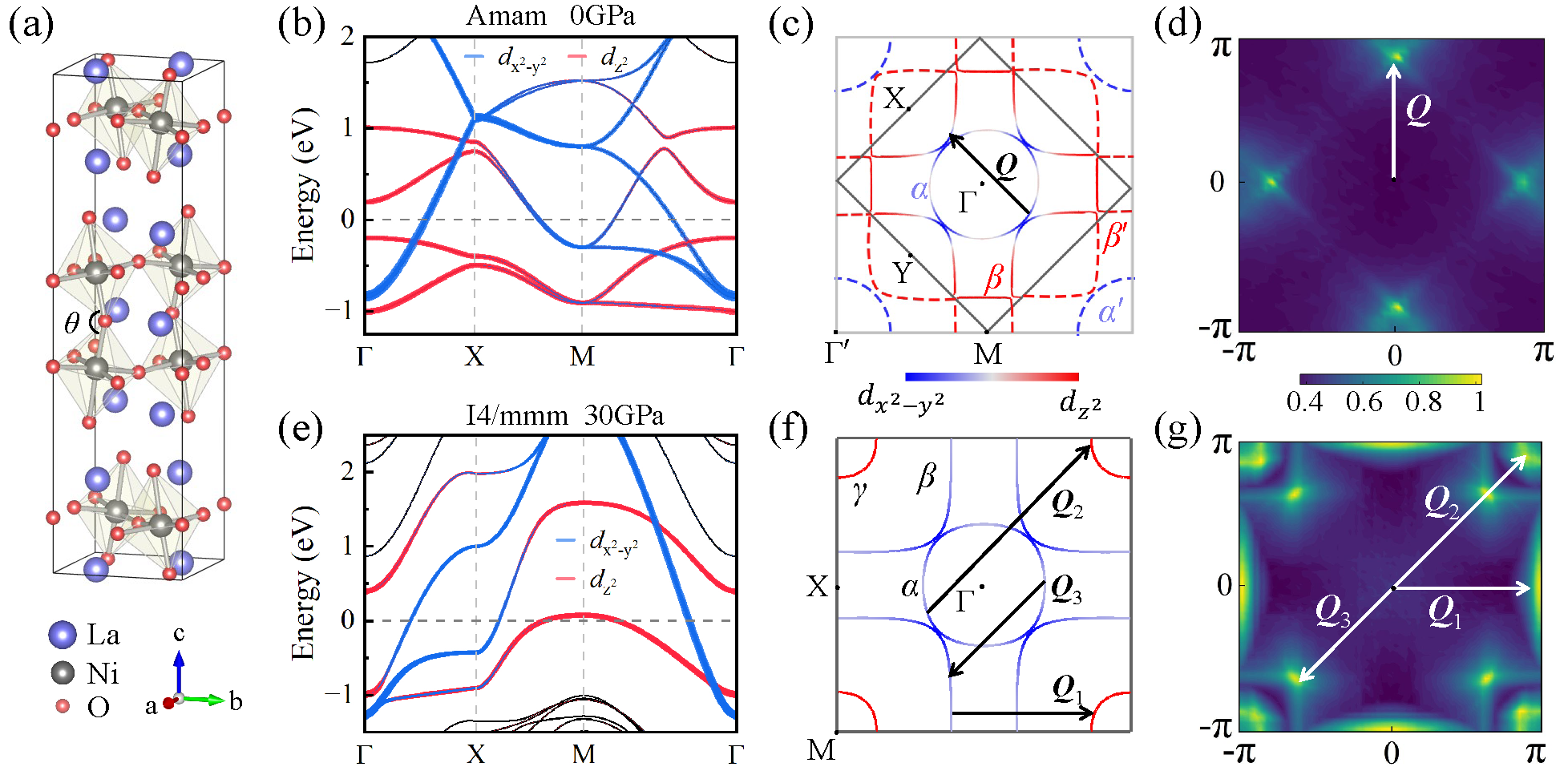}
    	\caption{Crystal and electronic structures, Fermi surfaces, and spin susceptibility of La$_3$Ni$_2$O$_7$ at ambient pressure and 30 GPa. (a) Amam crystal structure of La$_3$Ni$_2$O$_7$ at ambient pressure, with $\theta$ denoting the interlayer Ni--O--Ni bond angle. (b) and (e) Orbital-projected band structures of the Amam and I4/mmm phases, respectively. Blue and red line widths represent the Ni-$3d_{x^2-y^2}$ and Ni-$3d_{z^2}$ orbital weights. (c) and (f) Corresponding orbital-resolved Fermi surfaces projected onto the $k_x$--$k_y$ plane, with relevant nesting vectors indicated. In (c), the black diamond and gray square indicate the folded and unfolded BZ, respectively (the dashed $\alpha'$ and $\beta'$ are folded replicas of the $\alpha$ and $\beta$ sheets). (d) and (g) Static spin susceptibility $\chi_S(\bm{q})$ at ambient pressure and 30 GPa, respectively.}
    	\label{Figure/1.png}
    \end{figure*}

    In this work, we investigate the pressure-dependent electronic structure, spin response, and magnetic ground state of La$_3$Ni$_2$O$_7$, focusing on the interplay among its spin, charge, and lattice degrees of freedom. We establish a direct correspondence between the ambient-pressure FS nesting and the experimentally observed density-wave vector. At high pressure, the spin susceptibility arising from $\beta$--$\gamma$ nesting follows a pressure dependence closely paralleling the experimental $T_{\mathrm C}$, pointing to a possible role of nesting-enhanced spin fluctuations in superconducting pairing. Among the representative magnetic configurations considered, spin-charge-stripe states are energetically favored, with the $\bm{a}$-stripe state lying lowest in energy. It simultaneously develops inequivalent Ni moments, Ni--O bond-length disproportionation, and relaxes into the experimentally proposed polar Am2m structure. Together, these results establish a coherent microscopic picture in which strong spin-charge-lattice coupling stabilizes the low-pressure density-wave state.
 	
	\section{Methodology}
	First-principles calculations based on density functional theory (DFT) were carried out with the Vienna \textit{Ab initio} Simulation Package (VASP) \cite{Kresse1996,Blochl1994}. The generalized gradient approximation (GGA) in the Perdew--Burke--Ernzerhof (PBE) form was employed for the exchange-correlation functional. A fine $k$-point mesh with a resolution of 0.03~\AA$^{-1}$ was used for the self-consistent-field (SCF) calculations, whereas a denser mesh with a resolution of 0.007~\AA$^{-1}$ was employed for the FS calculations. The projector augmented-wave (PAW) method was applied with a plane-wave kinetic-energy cutoff of 520 eV. The phonon spectra were calculated using the density functional perturbation theory (DFPT) approach \cite{Baroni1987,Gonze1995a,Gonze1995b} and analyzed by the PHONOPY software \cite{Chaput2011,Togo2015}. To account for the correlation effects of the Ni-$3d$ electrons in La$_3$Ni$_2$O$_7$, the DFT+$U$ method was employed with an effective Hubbard $U_{\mathrm{eff}}=4$ eV, consistent with previous studies \cite{Zhang2024}.
    
    To investigate the microscopic origin of the magnetic instability, we calculated the static spin susceptibility $\chi_S(\bm{q},\omega=0)$ within a simplified random phase approximation (RPA) treatment as
    
    \begin{equation}
    \chi_S(\bm{q})=\frac{\chi_0(\bm{q})}{1-\chi_0(\bm{q})U},
    \label{eq:rpa}
    \end{equation}
    where $U$ denotes the effective interaction strength and $\chi_0(\bm{q})$ is the bare susceptibility \cite{Chan1973}. In the static limit and within the constant-matrix-element approximation, $\chi_0$ is evaluated as
    \begin{equation}
    \chi_0(\bm{q})=-\frac{1}{N}\sum_{\bm{k},m,n}\frac{f(\epsilon_{n,\bm{k}})-f(\epsilon_{m,\bm{k}+\bm{q}})}{\epsilon_{n,\bm{k}}-\epsilon_{m,\bm{k}+\bm{q}}},
    \label{eq:chi0}
    \end{equation}
   
    where $m$ and $n$ denote the band indices, $\epsilon_{n,\bm{k}}$ is the Kohn--Sham eigenvalue, and $f(\epsilon)$ is the Fermi--Dirac distribution function.
	
	\section{Results and discussion}

    \begin{figure*}
    	\centering
    	\includegraphics[width=0.8\linewidth]{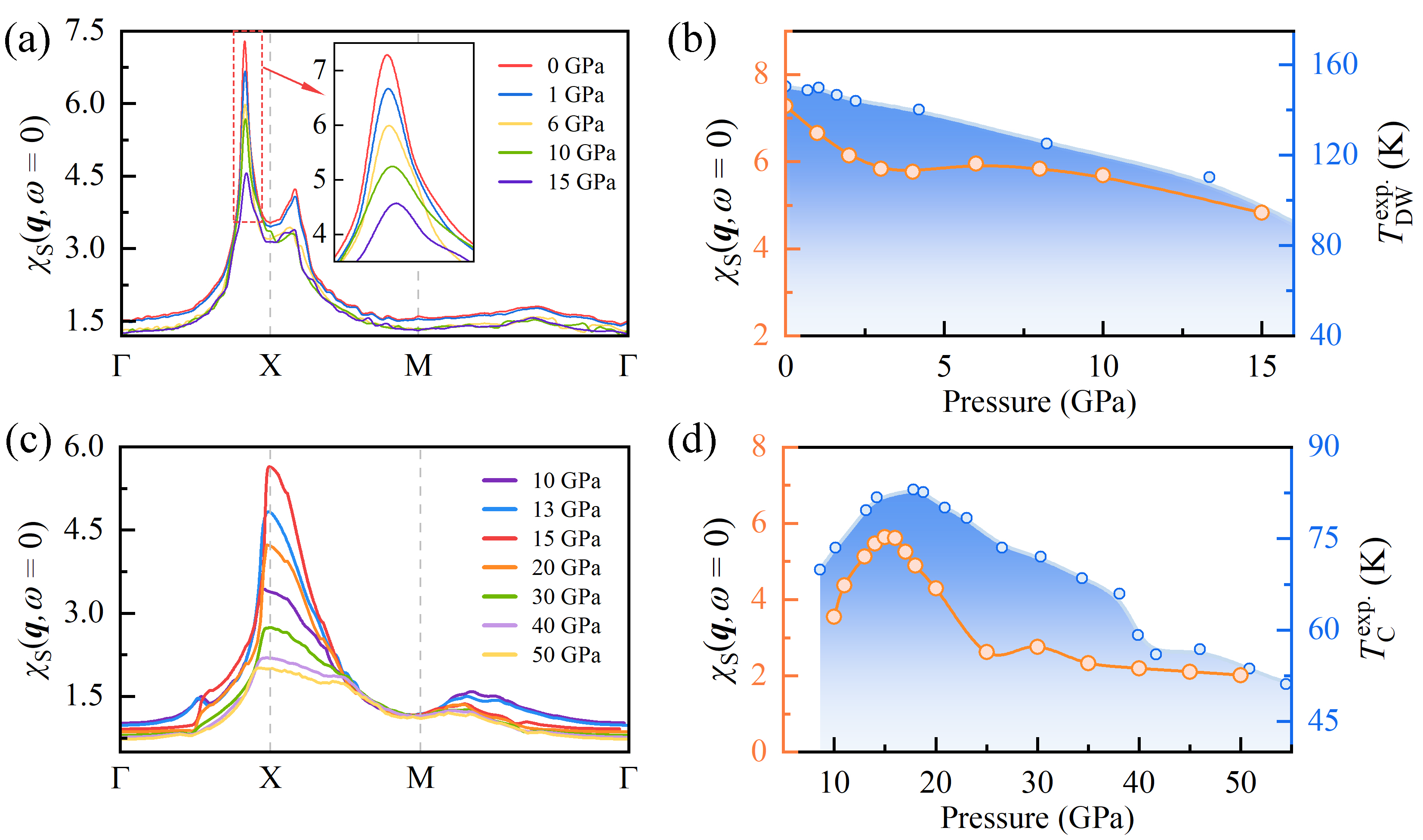}
    	\caption{Pressure evolution of static spin susceptibility $\chi_S(\bm{q})$ of La$_3$Ni$_2$O$_7$ and comparison with the experimental transition temperatures. (a) and (c) $\chi_S(\bm{q})$ along the $\Gamma$-$X$-$M$-$\Gamma$ path at selected pressures. The inset in (a) highlights the dominant peak near the $X$ point. In (c), the peak near the $X$ point arises primarily from nesting between the $\beta$ and $\gamma$ FS sheets, characterized by $\bm{Q}_1$. (b) and (d) Pressure dependence of the peak value of $\chi_S(\bm{q})$ near $X$ (orange), compared with the experimental density-wave-like transition temperature $T_{\mathrm{DW}}^{\mathrm{exp.}}$ \cite{Meng2024} and superconducting transition temperature $T_{\mathrm C}^{\mathrm{exp.}}$ \cite{Li2025NSR} (blue), respectively.}
    	\label{Figure/2.png}
    \end{figure*}
	
    At ambient pressure, La$_3$Ni$_2$O$_7$ is conventionally described by the orthorhombic Amam structure (space group No.~63) \cite{Chen2024PRL,Li2025NSR}, depicted in Fig.~\ref{Figure/1.png}. The structure comprises bilayers of corner-sharing NiO$_6$ octahedra separated along the crystallographic $c$ axis by fluorite-like La--O layers. The orthorhombic distortion breaks the in-plane fourfold rotation and makes the crystallographic $\bm{a}$ and $\bm{b}$ directions inequivalent, a feature directly relevant to the orientation of the stripe states discussed below.

    To uncover the microscopic origin of the ambient-pressure density-wave instability, we first examine how the low-energy electronic structure predisposes La$_3$Ni$_2$O$_7$ toward a particular ordering tendency. As shown in Fig.~\ref{Figure/1.png}(b), the electronic states near the Fermi level are dominated by the Ni-$3d_{x^2-y^2}$ and Ni-$3d_{z^2}$ orbitals, forming the multiband FS shown in Fig.~\ref{Figure/1.png}(c). Nearly parallel portions of the $\alpha$ and $\beta$ sheets are connected by a characteristic nesting vector $\bm{Q}$. Consistently, the static spin susceptibility in Fig.~\ref{Figure/1.png}(d) exhibits pronounced peaks near $\pm\bm{Q}\approx(0,0.87\pi)$ in the folded BZ. Upon unfolding, this wave vector maps onto a diagonal ordering vector close to the experimentally measured $\bm{Q}_{\mathrm{SDW}}\approx(\pi/2,\pm\pi/2)$ \cite{Gupta2025}.
    
    Having identified the dominant nesting-driven instability at ambient pressure, we now examine how the associated spin response evolves with pressure. As shown in Fig.~\ref{Figure/2.png}(a), the susceptibility peak near the $X$ point is progressively suppressed with increasing pressure, while its wave-vector position changes only slightly. To quantify this trend against experiment, Fig.~\ref{Figure/2.png}(b) plots the pressure dependence of the peak value alongside the measured density-wave transition temperature $T_{\mathrm{DW}}^{\mathrm{exp.}}$ \cite{Meng2024}. The two quantities exhibit a similar decreasing trend, indicating that the pressure-induced suppression of the density-wave state tracks the weakening of the nesting-enhanced spin response.
    
    Upon further compression, La$_3$Ni$_2$O$_7$ undergoes pronounced changes in both its crystal and electronic structures. In particular, the interlayer Ni--O--Ni linkage straightens, with the bond angle $\theta$ evolving from approximately $168^\circ$ at ambient pressure to $180^\circ$ in the high-pressure phase. The accompanying reconstruction of the bands near the Fermi level [Fig.~\ref{Figure/1.png}(e)] gives rise to a markedly different FS from that at ambient pressure. As shown in Fig.~\ref{Figure/1.png}(f), the reconstructed FS consists of the $\alpha$, $\beta$, and $\gamma$ sheets, with $\bm{Q}_1$, $\bm{Q}_2$, and $\bm{Q}_3$ marking characteristic nesting vectors connecting different FS sheets. The corresponding momentum-space distribution of the spin susceptibility is shown in Fig.~\ref{Figure/1.png}(g), where distinct response features associated with these nesting channels can be identified.

    Motivated by previous proposals that $\beta$--$\gamma$ nesting may be relevant to superconductivity \cite{Liu2025PRB,Zhang2024}, we focus on this specific channel and track the pressure evolution of its susceptibility peak. As shown in Fig.~\ref{Figure/2.png}(c), the $\chi_S(\bm{q})$ peak amplitude exhibits a nonmonotonic pressure dependence that closely follows that of the experimental $T_{\mathrm C}^{\mathrm{exp.}}$ [Fig.~\ref{Figure/2.png}(d)] \cite{Li2025NSR}. A similar correlation is found in rare-earth-substituted La$_3$Ni$_2$O$_7$ compounds at 30 GPa, where the $X$-point spin-susceptibility peak evolves with composition in agreement with the experimental $T_{\mathrm C}^{\mathrm{exp.}}$ trend, as detailed in Fig.~S1 of the Supplemental Material. Taken together, the pressure- and composition-dependent correlations indicate that spin fluctuations arising from $\beta$--$\gamma$ FS nesting may contribute to superconducting pairing, consistent with earlier proposals of $s\pm$-wave pairing \cite{Liu2023}.

    \begin{table}[t]
\caption{Relative energies and local Ni magnetic moments of the candidate magnetic states of La$_3$Ni$_2$O$_7$. Relative energies $\Delta E$ (meV/f.u.) are given with respect to the A-AFM state. Distinct moment values correspond to inequivalent Ni sites. NM, FM, A-AFM, G-AFM, DS, and SCS denote the nonmagnetic, ferromagnetic, A-type antiferromagnetic, G-type antiferromagnetic, double-stripe, and spin-charge-stripe configurations, respectively.}
\label{tab:magnetic}
\begin{ruledtabular}
\renewcommand{\arraystretch}{1.2}
\begin{tabular}{cccc}

\makecell{Magnetic\\configuration} &
\makecell{Stripe\\orientation} &
\makecell{$\Delta E$\\(meV/f.u.)} &
\makecell{Ni moments\\($\mu_B$)} \\
\colrule
NM    & --  & 1286.88 & 0 \\
FM    & --  & 20.55   & 1.45 \\
A-AFM & --  & 0.00    & 1.41 \\
G-AFM & --  & 219.54  & 1.33 \\
DS    & $\bm{a}$ & 61.42   & 1.33 \\
DS    & $\bm{b}$ & 68.04   & 1.34 \\
SCS   & $\bm{a}$ & -13.26  & 1.59/0.90 \\
SCS   & $\bm{b}$ & -9.45   & 1.59/0.90 \\

\end{tabular}
\end{ruledtabular}
\end{table}

    \begin{figure*}
    	\centering
    	\includegraphics[width=0.82\linewidth]{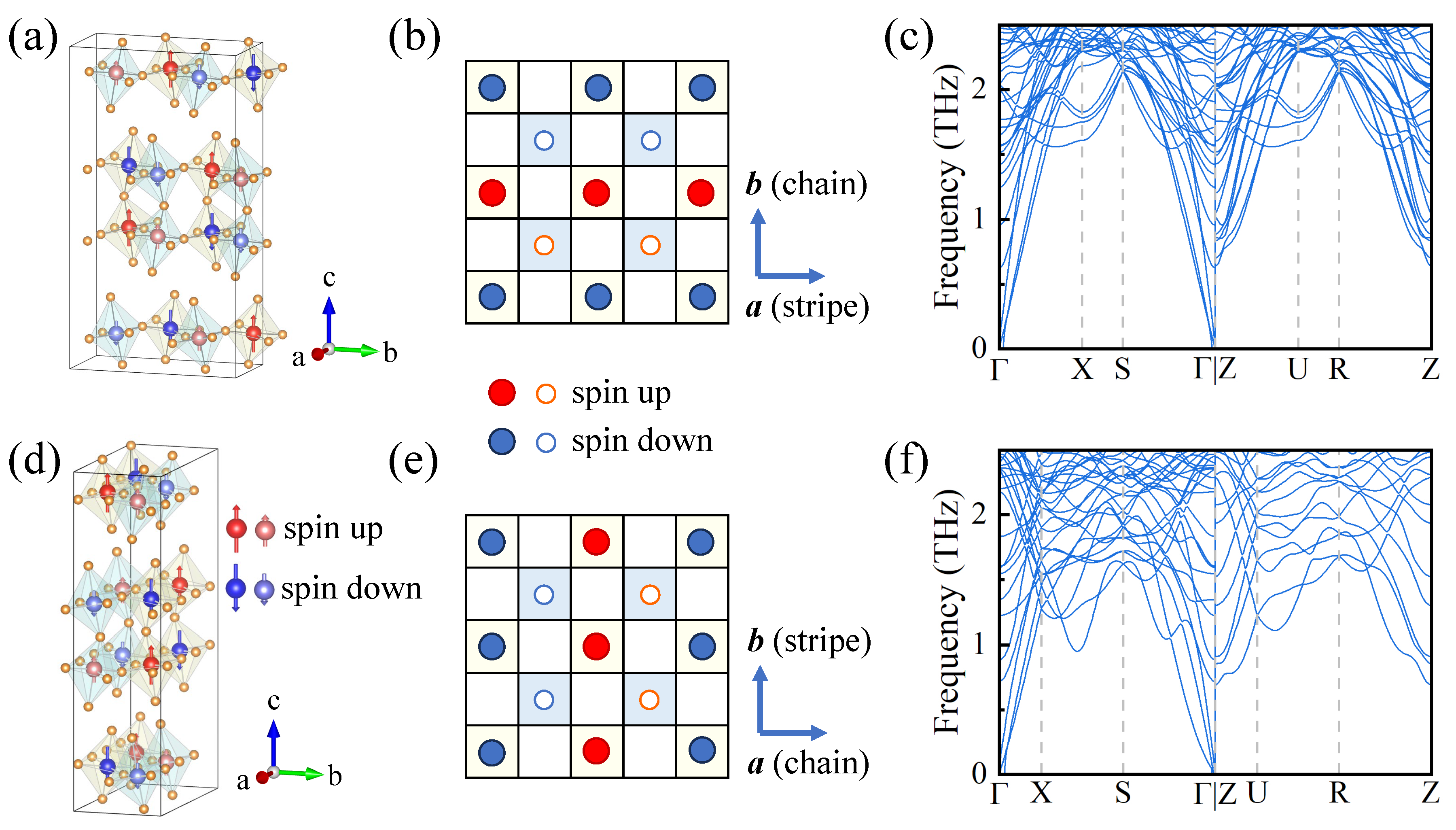}
    	\caption{Magnetic configurations and phonon dispersions of the $\bm{a}$- and $\bm{b}$-oriented spin-charge-stripe states in La$_3$Ni$_2$O$_7$. (a) and (d) Three-dimensional views of the relaxed $\bm{a}$-stripe and $\bm{b}$-stripe magnetic structures, respectively. Red and blue arrows denote spin-up and spin-down Ni moments, respectively. (b) and (e) Schematic top views of the corresponding spin arrangements in the NiO$_2$ plane. The circle color denotes the spin direction, while the circle size and color saturation distinguish Ni sites with larger and smaller moment magnitudes. (c) and (f) Low-frequency phonon dispersions of the $\bm{a}$-stripe and $\bm{b}$-stripe states, respectively, along the selected high-symmetry path.}
    	\label{Figure/3.png}
    \end{figure*}
    
    Having established the close connection between FS nesting and the pressure-dependent spin response, we now turn to the microscopic magnetic configuration underlying the ambient-pressure SDW state. We compare the relative energies and local Ni moments of a series of representative magnetic configurations, with distinct stripe orientations, summarized in Table~\ref{tab:magnetic}. Taking the A-AFM state, identified in previous calculations as a low-energy conventional magnetic configuration \cite{Yi2024}, as a benchmark, we find that both spin-charge-stripe states are further stabilized. In particular, the $\bm{a}$-stripe configuration has the lowest energy, whereas its $\bm{b}$-stripe state lies only 3.81 meV/f.u. higher, indicating close energetic competition between the two stripe orientations. In the double-stripe states, all Ni sites are initialized with the same moment magnitude, whereas the spin-charge-stripe states are constructed with inequivalent Ni moments. After full self-consistent relaxation, the magnetic-moment disproportionation remains pronounced, confirming the robustness of the spin-charge-stripe state against electronic relaxation.

    	\begin{figure*}
    	\centering
    	\includegraphics[width=0.9\linewidth]{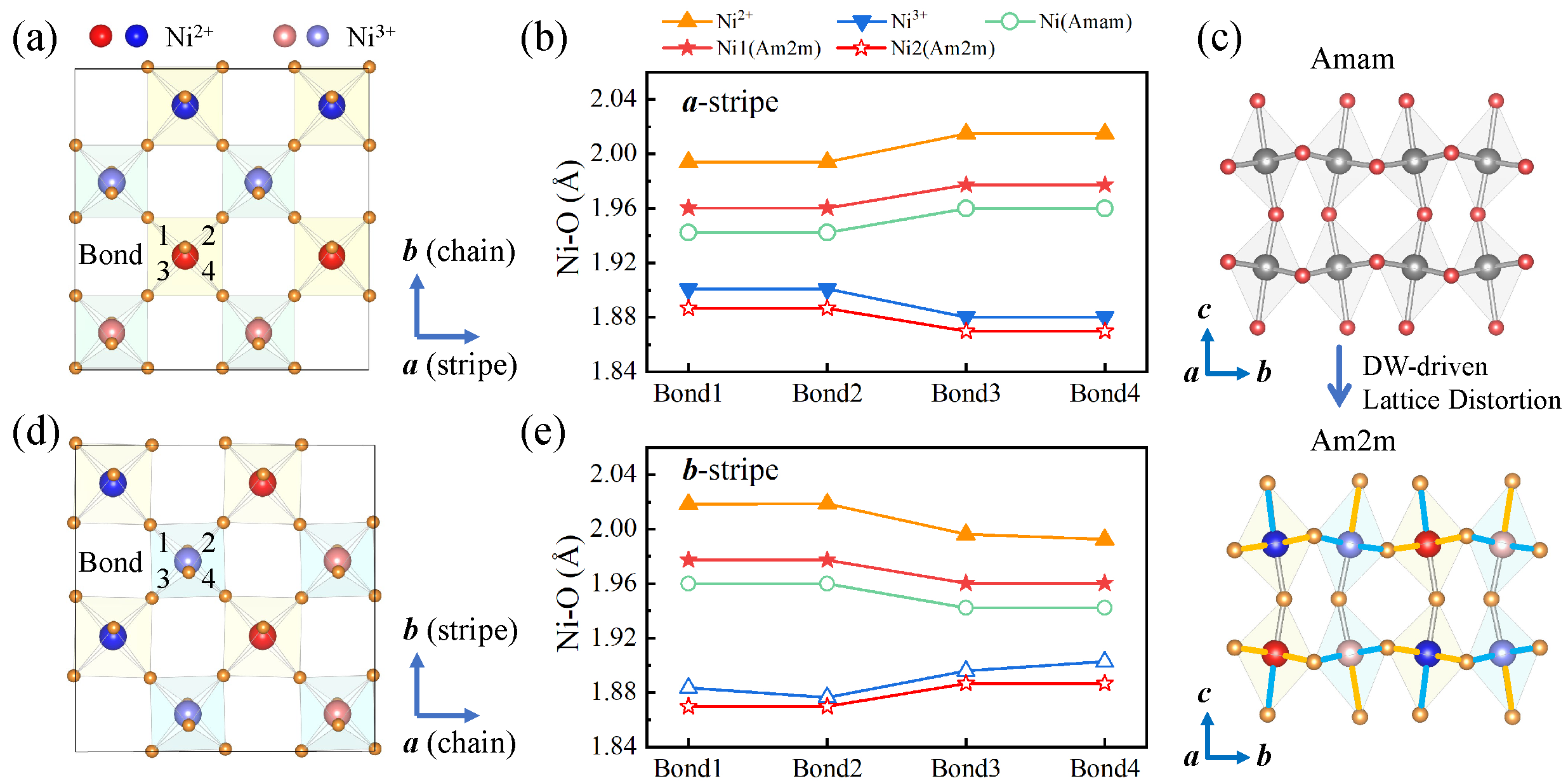}
    	\caption{Local Ni--O bond disproportionation and DW-driven symmetry lowering in La$_3$Ni$_2$O$_7$. (a) and (d) Top view of the relaxed $\bm{a}$-stripe and $\bm{b}$-stripe states, respectively. Red and blue indicate spin-up and spin-down moments, respectively, whereas the darker and lighter colors denote the Ni$^{2+}$-like and Ni$^{3+}$-like sites, respectively. The four in-plane Ni--O bonds are labeled Bond1--Bond4. (b) and (e) Corresponding Ni--O bond lengths, compared with those of the reference Amam and experimentally proposed Am2m structures. (c) Side views of the parent Amam structure and the relaxed $\bm{a}$-stripe structures, illustrating the spontaneous symmetry lowering to polar Am2m. Yellow and blue bonds indicate Ni--O bond elongation and contraction, respectively, relative to the parent Amam structure.}
    	\label{Figure/4.png}
    \end{figure*}
    
    To further characterize the two spin-charge-stripe states, we inspect their relaxed magnetic structures and phonon spectra, presented in Fig.~\ref{Figure/3.png}. The relaxed $\bm{a}$- and $\bm{b}$-oriented spin-charge-stripe configurations differ primarily in the stripe orientation relative to the orthorhombic $\bm{a}$ and $\bm{b}$ axes, following the convention of Ref.~\cite{Tokunaga2006}. Their in-plane magnetic arrangements, viewed along the crystallographic $c$ axis, are schematically illustrated in Figs.~\ref{Figure/3.png}(b) and \ref{Figure/3.png}(e). Importantly, neither configuration shows imaginary-frequency modes in phonon dispersions, establishing that both states are dynamically stable [Figs.~\ref{Figure/3.png}(c) and \ref{Figure/3.png}(f)]. Combined with the energy comparison in Table~\ref{tab:magnetic}, these results single out the $\bm{a}$-stripe state as the most favorable magnetic configuration among those considered.

    To examine whether the Ni-moment disproportionation is coupled to the lattice, we compare in Fig.~\ref{Figure/4.png} the relaxed Ni--O environments of the two spin-charge-stripe states with those of the reference Amam and Am2m structures. Figures~\ref{Figure/4.png}(b) and \ref{Figure/4.png}(e) compare the individual in-plane Ni--O bond lengths. In the reference Amam structure, octahedral tilting already gives rise to anisotropic Ni--O bond lengths within each NiO$_6$ octahedron, yet all Ni sites remain crystallographically equivalent. By contrast, stabilization of the spin-charge-stripe states produces pronounced site-dependent bond disproportionation: the larger-moment Ni sites exhibit longer Ni--O bonds, whereas the smaller-moment sites exhibit shorter ones. This systematic correlation reveals a strong coupling between the magnetic disproportionation and the local lattice response.

    Stoichiometric La$_3$Ni$_2$O$_7$ has a nominal average Ni valence of $+2.5$, whereas the recently proposed Am2m structure contains two inequivalent Ni sites with distinct Ni--O bonding environments and concomitant valence differentiation \cite{Misawa2026}. The concurrent magnetic-moment and bond-length disproportionation found here therefore suggests an accompanying charge redistribution, consistent with previous reports of site-selective valence differentiation with Ni$^{2+}$/Ni$^{3+}$-like character \cite{Chen2024PRL,Wang2024,Chen2025}. Notably, the $\bm{a}$-stripe state exhibits a pairwise bond-length relation, B1 = B2 and B3 = B4, closely resembling the local bonding pattern of the Am2m structure. Our results thus suggest a microscopic origin for the Ni-site differentiation associated with the Am2m structure.

    Symmetry analysis of the fully relaxed structures further uncovers a pronounced orientation-dependent lattice response. Starting from the centrosymmetric Amam structure, the $\bm{a}$-stripe state spontaneously relaxes into the polar Am2m structure (space group No.~38), whereas the $\bm{b}$-stripe state lowers the symmetry to monoclinic Cm (No.~8). Together with its lower total energy and dynamical stability, the $\bm{a}$-stripe state emerges as the most plausible magnetic ground state among those considered. Its relaxation into the experimentally proposed polar Am2m phase underscores the strong coupling between stripe magnetism and lattice-symmetry breaking. The competing $\bm{b}$-stripe state likewise shows that alternative magnetic orders pick out a different low-symmetry lattice response.
   
  \section{Conclusions}
	
   In summary, our calculations establish a microscopic connection among the electronic, magnetic, and structural instabilities of La$_3$Ni$_2$O$_7$. At low pressures, strong FS nesting produces a pronounced spin-susceptibility peak near the experimentally observed SDW ordering vector, which is progressively suppressed with pressure in parallel with the reduction of the density-wave transition temperature, $T_{\mathrm{DW}}$. In the high-pressure phase, FS reconstruction gives rise to pronounced $\beta$--$\gamma$ nesting, whose associated spin response exhibits a nonmonotonic pressure dependence that mirrors $T_{\mathrm C}$, indicating that nesting-enhanced spin fluctuations may contribute to superconducting pairing. Among the candidate magnetic states considered, the energetically favored spin-charge-stripe states are dynamically stable, with pronounced Ni-moment and Ni--O bond-length disproportionation. Most notably, the lowest-energy $\bm{a}$-stripe state spontaneously relaxes to the proposed Am2m structure, whereas the competing $\bm{b}$-stripe state adopts monoclinic Cm symmetry. Together, these results furnish a microscopic framework connecting stripe magnetism, local charge and bond disproportionation, and lattice symmetry lowering, underscoring the strong spin-charge-lattice coupling in the low-pressure density-wave state of La$_3$Ni$_2$O$_7$.

\begin{acknowledgements}
	This work was supported by the National Key Research and Development Program of China under Contract No.~2022YFA1403200. 
\end{acknowledgements}	
 
\bibliography{references}

\end{document}